\documentclass[sigconf]{acmart}
\usepackage{amsmath}
\usepackage{array}
\usepackage[caption=false,font=normalsize,labelfont=sf,textfont=sf]{subfig}
\usepackage{textcomp}
\usepackage{stfloats}
\usepackage{url}
\usepackage{verbatim}
\usepackage{threeparttable} 
\usepackage{graphicx}
\usepackage{tcolorbox}
\usepackage{multirow}
\usepackage{booktabs}
\usepackage{makecell}
\usepackage{tikz}
\usepackage{adjustbox}
\usepackage{xfp}
\usepackage{siunitx}
\usepackage{tabularx} 
\usepackage{listings}
\usepackage{comment}
\usepackage{balance}
\usepackage{diagbox}
\usepackage{svg}
\usepackage[table,xcdraw]{xcolor}
\usepackage{fontspec}  

\newcommand{\tool}{APTrans}

\usepackage[linesnumbered,ruled,vlined]{algorithm2e}
\usepackage{float}  
\SetKwComment{Comment}{$\blacktriangleright$~}{}        
\SetCommentSty{itshape\color{black}}                    
\SetEndCharOfAlgoLine{\relax}                           

\newcommand{\TriComment}[1]{\hfill{\footnotesize\textit{$\blacktriangleright$~#1}}}

 \usepackage{fontspec}

\definecolor{keywordcolor}{RGB}{166,38,164}
\definecolor{commentcolor}{RGB}{160,161,167}
\definecolor{stringcolor}{RGB}{255,140,0}
\definecolor{outputcolor}{RGB}{120,120,120} 

\lstdefinelanguage{CustomSQL}{
  morekeywords={
    CREATE, TABLE, INSERT, INTO, VALUES, SET, WITH, INT, PRIMARY, KEY, BEGIN, ROLLBACK, TRIGGER, ON, FOR, EACH, END, 
    REPEATABLE, READ, COMMITTED, START, COMMIT, SELECT, FROM, WHERE, UPDATE
  },
  sensitive=false,
  morecomment=[s]{/*}{*/},
  morestring=[b]',
}

\makeatletter
\renewcommand{\thelstnumber}{\@arabic\c@lstnumber.}
\makeatother

\AtBeginDocument{%
  }

\setcopyright{acmlicensed}
\copyrightyear{2026}
\acmYear{2026}
\acmDOI{XXXXXXX.XXXXXXX}
\acmConference[SIGMOD]{SIGMOD International Conference on Management of Data (2026)}{May 31--June 05, 2026}{Bengaluru, India}
\acmISBN{978-1-4503-XXXX-X/2026/06}

\begin{document}

\title{Anomaly Pattern-guided Transaction Bug Testing in Relational Databases}

\author{Shuang Liu, Huicong Xu, Xianyu Zhu, Qiyu Zhuang, Wei Lu, Xiaoyong Du}
\affiliation{%
  \institution{Renmin University of China}
}
\email{{shuang.liu, huicongxu, xianyuzhu, qyzhuang, lu-wei, duyong}@ruc.edu.cn}





\begin{abstract}
Concurrent transaction processing is a fundamental capability of Relational Database Management Systems (RDBMSs), widely utilized in applications requiring high levels of parallel user interaction, such as banking systems, e-commerce platforms, and telecommunications infrastructure. Isolation levels offer a configurable mechanism to manage the interaction between concurrent transactions, enabling varying degrees of consistency and performance trade-offs. These isolation guarantees are supported by all major RDBMSs. 
However, testing transaction behavior under different isolation levels remains a significant challenge due to two primary reasons. First, automatically generating test transactions that can effectively expose bugs in transaction handling logic is non-trivial, as such bugs are typically triggered under specific transactional constraints. Second, detecting logic anomalies in transaction outcomes is difficult because the correct execution results are often unknown for randomly generated transactions. 
To address these challenges, we propose an anomaly pattern-guided testing approach for uncovering transaction bugs in RDBMSs. Our solution tackles the first challenge by introducing a test case generation technique guided by predefined anomaly patterns, which increases the likelihood of exposing transactional bugs. For the second challenge, we present a two-phase detection process, involving explicit error detection and implicit error detection, to identify bugs in transaction execution. 
We have implemented our approach in a tool, \tool{}, and evaluated it on three widely-used RDBMSs: MySQL, MariaDB, and OceanBase. \tool{} successfully identified 13 previously unknown transaction-related bugs, 11 of which have been confirmed by the respective development teams.
 
\end{abstract}

\ccsdesc[500]{Do Not Use This Code~Generate the Correct Terms for Your Paper}
\ccsdesc[300]{Do Not Use This Code~Generate the Correct Terms for Your Paper}
\ccsdesc{Do Not Use This Code~Generate the Correct Terms for Your Paper}
\ccsdesc[100]{Do Not Use This Code~Generate the Correct Terms for Your Paper}


\maketitle

\section{Introduction}
\label{sec:introduction}


Relational Database Management Systems (RDBMSs) are foundational to a wide range of applications, particularly those demanding high levels of concurrency, such as telecommunications and financial services. A core feature of RDBMSs is their ability to support concurrent transaction processing, allowing multiple users to simultaneously access and manipulate shared data. However, this concurrency introduces significant challenges in ensuring data consistency and correctness. To mitigate these challenges, the SQL standard~\cite{melton2016iso} defines a hierarchy of isolation levels—Read Uncommitted, Read Committed, Repeatable Read, and Serializable—each offering a distinct trade-off between performance and isolation guarantees. Thus, ensuring the correctness of concurrent transaction executions requires comprehensive testing that aligns with the formal semantics of each isolation level and the anomalies they may permit.

In practice, database systems employ various concurrency control mechanisms—such as pessimistic locking~\cite{thomasian1993two}, optimistic concurrency control (OCC)~\cite{kung1981optimistic}, and multiversion concurrency control (MVCC)~\cite{bernstein1983multiversion}—to implement transaction isolation as defined by the SQL standard. To optimize performance, these mechanisms are often supplemented with system-specific enhancements. However, the added complexity of these optimizations can introduce subtle bugs, which may lead to violations of the isolation guarantees. Ensuring correctness in the presence of these optimizations remains a critical challenge in the development and testing of database management systems (DBMSs).



\begin{figure}[t] 
\centering
\begin{lstlisting}
/*init*/ CREATE TABLE t1 (c0 INT);
/*init*/ CREATE TABLE t2 (c0 INT PRIMARY KEY AUTO_INCREMENT, c1 INT);
/*init*/ INSERT INTO t1(c0) VALUES (1),(2),(3);
/*txn1*/ SET SESSION TRANSACTION ISOLATION LEVEL SERIALIZABLE;
/*txn2*/ SET SESSION TRANSACTION ISOLATION LEVEL SERIALIZABLE;
/*txn1*/ BEGIN;
/*txn2*/ BEGIN;
/*txn1*/ INSERT INTO t2(c1) VALUES(1);
/*txn2*/ INSERT INTO t2(c1) VALUES(2), (3);
/*txn2*/ COMMIT;
/*txn1*/ UPDATE t1 SET c0 = 10 WHERE c0 in (SELECT c1 FROM (SELECT * FROM t2 ORDER BY c0 LIMIT 2) AS t);
/*txn1*/ COMMIT;
/*Finally database state*/
SELECT * FROM t1; @−[( 10, 10, 3 )]@ 
\end{lstlisting}
\caption{Motivation example: a logic bug detected by \tool{} in Mariadb (Bug\#36330)} 
\label{fig:motivation} 
\end{figure}
Figure~\ref{fig:motivation} illustrates a transaction bug at the Serializable isolation level in MariaDB~\cite{MariaDB}. In a high-demand concert ticket-booking system, two agents simultaneously process the remaining seats. Agent A ($txn1$) first reserves seat 1 for a customer. Meanwhile, Agent B (Transaction $txn2$) reserves seats 2 and 3 for another customer and completes the booking. When Agent A attempts to reserve one additional seat for their original customer by querying the system for the two lowest-numbered available seats, the serializable isolation should logically show Agent A only their originally reserved seat 1. However, due to a serialization anomaly, the system incorrectly assigns seat 2 to Agent A's customer as well, resulting in a double-booking scenario where two customers are assigned the same seat, creating a conflict that requires manual resolution. 

It is hard to detect the bug in Figure~\ref{fig:motivation} due to two main reasons. First, it is hard to trigger this bug as a certain transaction pattern, involving statements, execution order and shared data, is required in the test case (e.g., $txn1$ and $txn2$  start first, then $txn2$ inserts data into the shared table and commits, $txn1$ inserts into the same table, updates according to the insert and commit). Second, the bug does not trigger any explicit error message, and we need to uncover it by checking the database status after each statement. It is difficult to build an oracle to detect incorrect database status, as it is hard to determine the correct result from a single random transaction. 


Existing approaches for testing transaction anomalies, such as Troc~\cite{dou2023detecting} and TxCheck~\cite{jiang2023detecting}, typically generate test transactions by randomly selecting SQL statements from a predefined set. This makes it challenging to synthesize transactions like the one shown in Example~\ref{fig:motivation}, which involve a pattern with multiple constraints and complex dependencies. Specifically, in this example, not only must the SQL statements and schedule be deterministic, but it is also essential that different SQL statements access the same data items—an aspect that is difficult to achieve with randomly generated methods. 
Troc~\cite{dou2023detecting} adopts a differential testing strategy by constructing a view for each SQL statement and comparing the results of concurrent transaction executions against these view-based outputs. However, the process of constructing these views introduces considerable overhead, requiring the implementation of a parallel transaction execution engine. Furthermore, due to differences in implementation semantics, the differential approach often yields a high rate of false positives (as discussed in Section~\ref{sec:exp-compare}). 
TxCheck~\cite{jiang2023detecting}, on the other hand, attempts to decouple a transaction into independent and dependent SQL statements. It then generates semantically equivalent variants by reordering the independent statements while preserving the order of the dependent ones. The results of the two transaction executions are compared to identify inconsistencies. 
TxCheck cannot solve cases where transactions exhibit cyclic dependencies, such as the (\(W_1[x_1]\)\(W_2[y_1]\)\(C_2\)\(R_1[y_1]\)\(C_1\)) cycle in Example~\ref{fig:motivation}, which significantly limits its applicability to more complex transactional workloads. ELLE~\cite{kingsbury2020elle}, a transaction anomaly checker, detects anomalies by identifying cycles in the transaction execution history. It is unable to detect bugs that do not show cycles in the execution history, such as the dirty write anomaly in Table~\ref{tab:common_pattern}. 

In general, testing transaction bugs in RDBMSs raises two challenges. 
First, generating effective test cases capable of triggering transaction bugs is inherently difficult. Unlike traditional SQL functionality testing, transaction testing must account for complex interactions involving specific SQL statement types, concurrent data access patterns, and precise interleavings or scheduling orders. Bug-triggering transactions often conform to subtle and non-trivial execution patterns, making automated test generation a non-trivial task. 

%
Second, accurately detecting transaction bugs is equally challenging, particularly when dealing with silent bugs—those that do not result in explicit errors or exceptions. For automatically generated transactions, establishing the correct expected behavior is difficult, as it requires a deep understanding of both the intended application semantics and the database’s concurrency control mechanisms. 

To address the first challenge, we propose an anomaly-pattern-guided transaction generation approach. This method leverages anomaly transaction patterns—a high-level abstraction of known anomalous behaviors—to guide the synthesis of transactional workloads. The core idea is to instantiate these anomaly patterns using diverse SQL statements and database schemas. By doing so, we aim to exercise different concurrency control mechanisms and expose latent bugs in their implementation logic. 
To address the second challenge, we introduce a two-phase checking mechanism for detecting silent bugs. In the first phase, conduct explicit error checking by monitoring the database crash reports and assertion failures, a bug is reported on countering such cases. Otherwise, \tool{} invoke the second phase, where it analyzes the execution history of input transactions to identify potential anomaly patterns. If such a pattern is detected, it is reported as a bug, since these patterns should not occur under correct concurrency control implementations. Because our patterns are modeled based on data anomalies, the bugs we report are directly tied to specific data anomalies, making our detection method  accurate and interpretable.

We implemented our approach as an anomaly pattern-guided transaction bug testing tool,  \tool. We evaluated \tool{} on three widely-used and mature DBMSs: MySQL~\cite{Mysql}, MariaDB~\cite{MariaDB}, and OceanBase~\cite{OceanBase}. In total, \tool{} discovered 13 unique transaction bugs, of which 11 have been confirmed by the respective development teams and 10 are reported for the first time. Comparative experiments with state-of-the-art transaction bug testing tools demonstrate that \tool{} outperforms existing approaches in both bug-triggering effectiveness and bug detection accuracy. We have made \tool{} publicly available at \url{https://github.com/Paper-code-sigmod/APTrans}.

In summary, we make the following contributions. 
\begin{itemize}
    \item We propose a novel approach for testing transaction bugs by leveraging transaction anomaly patterns for both test case generation and result validation, enabling effective bug triggering and accurate bug detection.
    \item We conducted extensive experiments on three popular RDBMSs, uncovering 13 unique transaction bugs over a period of three months, 11 of which have been confirmed by the respective developers. 
    \item We implemented our approach in an open source tool, \tool, to encourage adoption and further research in the field. 
\end{itemize}
\section{Preliminary}
\label{sec:pre}

\subsection{Database Transactions and Isolation level}
 
Database transactions are fundamental for maintaining data integrity and consistency within a database management system, particularly in environments where multiple users or processes access the database concurrently. A transaction represents a logical unit of work composed of one or more database operations, all of which must be executed in a manner that ensures the database transits from one consistent state to another. Database transactions adhere to the ACID (Atomicity, Consistency, Isolation, and Durability) properties, which are essential for systems with stringent data integrity and reliability requirements, such as banking systems, e-commerce platforms. 

The isolation level determines the degree to which transactions are isolated from one another in a concurrent environment. It controls the visibility of one transaction’s operations to others, thereby striking a balance between performance and consistency. According to the SQL standard~\cite{melton2016iso}, there are four isolation levels——READ Uncommitted, Read Committed, Repeatable Read, and Serializable. The transaction isolation level defines the types of phenomena that may occur during the execution of concurrent transactions. The following phenomena are possible:
\begin{itemize}
    \item \textbf{Dirty Read (P1):} Transaction \( T_1 \) modifies a row, and  transaction \( T_2 \) reads that row before \( T_1 \) performs a \texttt{COMMIT}. If \( T_1 \) later performs a \texttt{ROLLBACK}, then \( T_2 \) read a row that was never committed. 
    \item \textbf{Non-repeatable Read (P2):}  Transaction \( T_1 \) reads a row, then  transaction \( T_2 \) modifies or deletes that row and commits. If \( T_1 \) later attempts to reread the row, it may obtain a different value than its initial read. 
    \item \textbf{Phantom (P3):} Transaction \( T_1 \) reads the set of rows \( N \) that satisfy a specific search condition. Transaction \( T_2 \) then executes statements that insert/delete one or more rows that satisfy the same search condition. If \( T_1 \) subsequently repeats the initial read with the same search condition, it may obtain a different set of rows. 
\end{itemize}
%
%
\begin{table}[t]
\centering
\begin{threeparttable}
\caption{Isolation level and the allowed phenomena}
\setlength{\tabcolsep}{10pt}
\begin{tabular}{llll}
\hline
\textbf{Isolation Level} & \textbf{P1} & \textbf{P2} & \textbf{P3} \\ \hline
Read Uncommitted (RU)   & P           & P           & P           \\
Read Committed (RC)      & NP          & P           & P           \\
Repeatable Read (RR)     & NP          & NP          & P           \\
Serializable (SER)       & NP          & NP          & NP          \\ \hline
\end{tabular}
\begin{tablenotes}
\small
\item \textbf{P} = Possible, \textbf{NP} = Not Possible.
\end{tablenotes}
\label{tab:phenomenon_Isolation}
\end{threeparttable}
\end{table}

The four isolation levels allow for different phenomena, reflecting the varying degrees of data integrity they offer. The correspondence of isolation levels and the phenomena allowed is shown in Table~\ref{tab:phenomenon_Isolation}. 

\subsection{Anomaly patterns}
\label{sec:dependency}



Adya~\cite{adya2000generalized} introduced the concept of transaction dependencies and dependency cycle to define data anomalies, and redefined isolation levels into PL-0, PL-1, PL-2, PL-2.99, and PL-3. According to Adya, different isolation levels prevent different anomalies. For example, PL-1 avoids rings formed by write-write dependencies. The dependency definitions are as follows, where \( T_i \prec T_j \) means that transaction \( T_i \) precedes transaction \( T_j \) in time:
\begin{itemize}
    \item \textbf{Write-write dependency (ww):} A transaction \( T_j \) directly ww-depends on \( T_i \) if \( T_i \) installs a version \( x_i \) of \( x \), and \( T_j \) installs the next version \( x_j \) of \( x \), with \( T_i \prec T_j \).
    \item \textbf{Write-read dependency (wr):} A transaction \( T_j \) directly wr-depends on \( T_i \) if \( T_i \) installs a version \( x_i \) and \( T_j \) reads \( x_i \).
    \item \textbf{Read-write dependency (rw):} A transaction \( T_j \) directly rw-depends on \( T_i \) if \( T_i \) reads version \( x_i \) of \( x \), and \( T_j \) installs the next version of \( x \), with \( T_i \prec T_j \).
\end{itemize}


A recent work by Li et al.~\cite{HaixiangLi} further extended the definition of data anomalies, categorizing them into 29 specific anomaly patterns as defined in~\ref{def:pattern}.

\begin{table}[t]
\scriptsize
\caption{Pattern definitions for common anomalies and the isolation levels that prevent them}
\label{tab:common_pattern}

\setlength{\tabcolsep}{8pt}
\begin{tabular}{m{2cm}m{3.5cm}m{1.5cm}}
\hline
\textbf{Anomaly} & \textbf{Pattern}  & \textbf{Disallowed in Isolation Levels}\tabularnewline \hline

 Dirty Write 
& $W_1[x_1]W_2[x_2]C_1C_2$
&  \textbf{RU}, \textbf{RC}, \textbf{RR}, \textbf{SER}\tabularnewline

 Dirty Read 
&$W_1[x_1]R_2[x_2]A_1C_2$
&  \textbf{RC}, \textbf{RR}, \textbf{SER}\tabularnewline

 Lost Update 
&$R_1[x_0]W_2[x_1]C_2W_1[x_2]C_1$
&  \textbf{RR}, \textbf{SER}\tabularnewline

 Non-Repeatable Read 
&$R_1[x_0]W_2[x_1]C_2R_1[x_1]C_1$
&  \textbf{RR}, \textbf{SER}\tabularnewline

 Write Skew 
&$R_1[x_0]W_2[x_1]W_2[y_1]C_2R_1[y_1]C_1$
&  \textbf{SER}\tabularnewline

 Read Skew 
&$R_1[x_0]R_2[y_0]W_1[y_1]W_2[x_1]C_1C_2$
&  \textbf{SER}\tabularnewline
\hline

\end{tabular}
\end{table}

\begin{definition}[Anomaly Pattern]
\label{def:pattern}
An anomaly pattern is defined as a sequence of operations, where each operation is annotated with its type (R/W/C), the transaction it belongs to, the target data item and its associated version. 
\end{definition}
For instance, the lost update anomaly is formalized with the anomaly pattern $R_1[x_0]W_2[x_1]C_2W_1[x_2]C_1$, 
Where $R_i,W_i$ and $C_i$ represent the read, write, and commit operations of transaction $T_i$;  $x_0,x_1$ and $x_2$ denote successive versions of data item $x$, with version numbers assigned based on write order (e.g., $x_0$ is the initial state, $x_1$ and $x_2$ are updated subsequently in order). 
This pattern contains two dependencies, $T_1 \xrightarrow{rw[x]} T_2$ and $T_2 \xrightarrow{ww[x]} T_1$, in Adya's formatting.

 Li's work~\cite{HaixiangLi} summarized 29 anomaly patterns. We selected representative ones as shown in Table~\ref{tab:common_pattern}, which contains transaction anomalies, along with their corresponding anomaly patterns and the isolation levels in which these anomalies are prohibited.
\section{Methodology}
\label{sec:method}

In this section, we provide an overview of our approach, followed by a detailed description of each component in our approach, i.e., database and SQL statement generation, pattern-guided transaction generation, as well as transaction execution and bug detection.

\subsection{Overview}



\begin{figure*}[t] 
\centering 
\includegraphics[width=\textwidth]{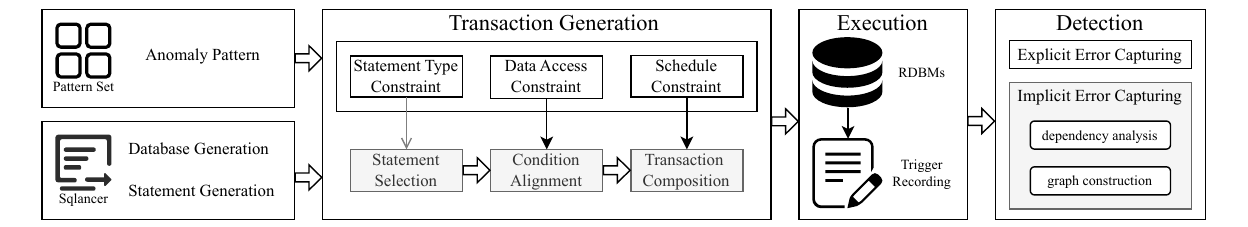} 
\caption{Overview of \tool} 
\label{fig:architect} 
\end{figure*}


To enable efficient transaction generation and precise bug detection, we propose \tool, a database transaction testing framework guided by anomaly patterns. The primary objective of \tool{} is to generate transactions that are highly effective at triggering bugs and to accurately detect transactional bugs. To this end, \tool{} introduces two key innovational: the anomaly pattern-guided transaction generation method designed to improve the bug-triggering ability of the generated transactions and a two-phase bug detection mechanism designed to identify and validate transaction-related bugs with high accuracy.
Figure~\ref{fig:architect} illustrates the overall architecture of \tool, which consists of four modules: database and SQL statement generation, transaction generation guided by the anomaly pattern, transaction execution, and transaction bug detection.

\subsection{Database and SQL statement Generation}
As the initial step of automated transaction generation, we first create the underlying databases on which the transactions will operate, along with the basic SQL statements that compose each transaction. We develop our database and SQL statement generation method based on SQLancer~\cite{SQLancer} and extend it to support table join operations for both database generation and SQL statement generation.  

\subsubsection{Database Generation}
In database transaction testing, the quality of the database structure and data initialization is critical to the validity of test cases. To simulate real-world scenarios, which usually involve multiple tables connected with join keys, we extend SQLancer to support table join operations for both database generation and SQL statement generation. 
Moreover, to support the transaction bug detection, which requires checking on the recorded data and the corresponding version, we introduced a unique row ID and a version column in the generated table. 

\vspace{1mm}
\noindent\textbf{Table Generation}. The base version of SQLancer supports the generation of isolated tables and the basic \texttt{CREATE TABLE} statements. It randomly selects column types such as \texttt{INT}, \texttt{TEXT}, and \texttt{BOOLEAN}, and applies common constraints such as \texttt{PRIMARY KEY}, \texttt{NOT NULL}, and \texttt{UNIQUE}, in the generated table creation statement. 
We extend the table generation functionality to support multi-table generation by adding foreign key constraints based on candidate key matching. When generating a foreign key constraint for a column in a table, we scan all existing \texttt{PRIMARY KEY} or \texttt{UNIQUE} columns in other tables and select candidate columns with matching data types to establish reference relationships. 


\vspace{1mm}
\noindent\textbf{Data Generation}. SQLancer generates random data by creating \texttt{INSERT} statements for tables. However, the random generation approach cannot guarantee the generated data satisfying the foreign key constraints we have implemented, as foreign keys require references to already existing values, and randomly generated values may not be present in the referenced columns. 
To address this issue, we introduced a data logging feature that records the primary key values of all inserted data. When generating data with foreign key constraints, we select existing values from the recorded data to ensure both data consistency and referential integrity.


\vspace{1mm}
\noindent\textbf{Unique Row ID and Version Column}. To enable traceability and version control of transaction operations, we introduce two special columns in each table: the \texttt{ROW ID} column (unique row identifier) and the \texttt{Version} column. The ROW ID column is an integer column that increments by one for each inserted record. It serves as a globally unique identifier to determine whether different transactions have accessed the same data item and forms the basis for constructing read/write dependency. The Version column is an integer column that records the version number of the current record, incrementing by one each time the record is modified. This column allows us to identify whether a transaction has read an old version of a data item, assisting in detecting  read/write version conflicts. Note that these two columns do not affect the actual execution semantics of the transaction; they are solely used for dependency reconstruction and anomaly detection within the testing framework.

\subsubsection{Statement Generation}
SQL statement generation is fundamental for testing transaction behavior, as its quality directly impacts the establishment of transaction dependencies and the effectiveness of subsequent bug detection. Building upon the SQLancer framework, we have extended the SQL statement generation to include JOIN operations and type constraints, thereby producing a more diverse set of SQL statements while ensuring the semantic correctness of the SQL statements.

\vspace{1mm}
\noindent\textbf{SQL Statement Generation}. We use SQLancer to generate a large number of random SQL statements for transaction generation. However, SQLancer suffers from numerous semantic errors, most of which are caused by type mismatches in expressions. To address this issue, we introduced type constraint checking during expression generation to ensure that the generated expressions conform to the required expression type.

\vspace{1mm}
\noindent\textbf{JOIN Statement Generation. } 
To support join statement on multiple tables in the database, we have extended the generation of SQL statements to include JOIN operations. Specifically, we generate four types of JOIN operations: \texttt{INNER JOIN}, \texttt{LEFT JOIN}, \texttt{RIGHT JOIN}, and \texttt{CROSS JOIN}. The core of generating the JOIN Statement lies in identifying columns for connection across different tables. We begin by scanning all columns in the current table. If we find a column with a foreign key constraint, we connect it with the corresponding primary key column. If a primary key column is found, we then search through all other tables to identify any foreign key referencing this column, establishing a connection if found. In other cases, we scan all tables and connect columns of the same type. Finally, we randomly select one of the JOIN methods for the identified columns and generate the corresponding JOIN statement. We skip the JOIN statements generation if no suitable columns are found for the given database. 


\subsection{Pattern-guided Transaction Generation}
\label{sec:transgen}

To enhance the bug triggering ability of the generated transaction, we propose a anomaly pattern-guided transaction generation method. We adopt the pattern formatting in Definition~\ref{def:pattern} and extended the existing anomaly patterns by Li et al.~\cite{HaixiangLi}. In total, we have identified 40 new anomaly patterns, detailed in the Table~\ref{tab:newpattern}. 
We have systematically enumerated all possible anomaly patterns involving two transactions and two shared variables. 
Our framework is designed with extensibility in mind, enabling seamless integration of additional anomaly patterns as needed.

\begin{table}[t] 
\footnotesize
\centering
\caption{Anomaly Patterns Extended by \tool}
\setlength{\tabcolsep}{5pt} 
\begin{tabular}{m{0.5cm}m{4cm}m{3cm}}
\hline
\textbf{ID}  & \textbf{Pattern} & \textbf{Disallowed in isolation levels} \\ \hline

1    & $W_1[x_1]R_2[x_1]C_1C_2$      & RC, RR, SER  \\
2    & $W_1[x_1]C_1R_2[x_1]C_2$      & RR           \\ 
3    & $W_1[x_1]R_2[x_1]C_2W_1[x_2]C_1$ & RC, RR, SER \\

4   & $W_1[x_1]W_2[y_1]W_2[x_2]W_1[y_2]C_1C_2$  & ALL       \\
5   & $W_1[x_1]W_2[y_1]W_1[y_2]W_2[x_2]C_1C_2$  & ALL       \\
6   & $W_1[x_1]W_2[x_2]C_2W_2[y_2]W_1[y_2]C_1$  & ALL       \\
7   & $W_1[x_1]W_2[y_1]W_2[x_2]C_2W_1[x_2]C_1$  & ALL       \\

8    & $W_1[x_1]W_2[y_1]W_2[x_2]R_1[y_1]C_1C_2$  & ALL       \\
9   & $W_1[x_1]W_2[y_1]R_1[y_1]W_2[x_2]C_1C_2$  & ALL       \\
10   & $W_1[x_1]W_2[y_1]W_2[x_2]C_2R_1[y_1]C_1$  & ALL       \\
11   & $W_1[x_1]W_2[y_1]R_1[y_1]C_1W_2[x_2]C_2$  & RC, RR, SER       \\

12   & $W_1[x_1]R_2[y_0]W_2[x_2]W_1[y_1]C_1C_2$  & ALL       \\
13   & $W_1[x_1]R_2[y_0]W_1[y_1]W_2[x_2]C_1C_2$  & ALL       \\
14   & $W_1[x_1]R_2[y_0]W_2[x_2]C_2W_1[y_1]C_1$  & ALL       \\
15   & $W_1[x_1]R_2[y_0]W_1[y_1]C_1W_2[x_2]C_2$  & RR, SER       \\ 

16   & $W_1[x_1]W_2[y_1]R_2[x_1]W_1[y_2]C_1C_2$  & ALL       \\
17   & $W_1[x_1]W_2[y_1]W_1[y_2]R_2[x_1]C_1C_2$  & ALL       \\
18   & $W_1[x_1]W_2[y_1]R_2[x_1]C_2W_1[y_2]C_1$  & RC, RR, SER       \\
19     & $W_1[x_1]W_2[y_1]W_1[y_2]C_1R_2[x_1]C_2$  & ALL       \\ 

20     & $W_1[x_1]W_2[y_1]R_2[x_1]R_1[y_1]C_1C_2$  & RC, RR, SER       \\
21     & $W_1[x_1]W_2[y_1]R_1[y_1]R_2[x_1]C_1C_2$  & RC, RR, SER       \\
22     & $W_1[x_1]W_2[y_1]R_2[x_1]C_2R_1[y_1]C_1$  & RC, RR, SER       \\
23     & $W_1[x_1]W_2[y_1]R_1[y_1]C_1R_2[x_1]C_2$  & RC, RR, SER       \\ 

24     & $W_1[x_1]R_2[y_0]R_2[x_1]W_1[y_1]C_1C_2$  & RC, RR, SER       \\
25     & $W_1[x_1]R_2[y_0]W_1[y_1]R_2[x_1]C_1C_2$  & RC, RR, SER       \\
26     & $W_1[x_1]R_2[y_0]R_2[x_1]C_2W_1[y_1]C_1$  & RC, RR, SER       \\
27     & $W_1[x_1]R_2[y_0]W_1[y_1]C_1R_2[x_1]C_2$  & RR, SER       \\ 

28     & $R_1[x_0]W_2[y_1]W_2[x_1]W_1[y_2]C_1C_2$  & RC, RR, SER       \\
29     & $R_1[x_0]W_2[y_1]W_1[y_2]W_2[x_1]C_1C_2$  & RC, RR, SER       \\
30     & $R_1[x_0]W_2[x_1]C_2W_2[y_1]W_1[y_2]C_1$  & RR, SER        \\
31     & $R_1[x_0]W_2[y_1]W_2[x_1]C_2W_1[x_1]C_1$  & ALL       \\ 

32     & $R_1[x_0]W_2[y_1]W_2[x_1]R_1[y_1]C_1C_2$  & RC, RR, SER       \\
33     & $R_1[x_0]W_2[y_1]R_1[y_1]W_2[x_1]C_1C_2$  & RC, RR, SER       \\
34     & $R_1[x_0]W_2[y_1]W_2[x_1]C_2R_1[y_1]C_1$  & SER        \\
35     & $R_1[x_0]W_2[y_1]R_1[y_1]C_1W_2[x_1]C_2$  & RC, RR, SER       \\ 

36       & $W_1[x_1]W_2[y_1]C_2R_1[y_1]C_1$        & SER      \\

37     & $R_1[x_0]R_2[y_0]W_2[x_1]W_1[y_1]C_1C_2$  & SER       \\
38     & $R_1[x_0]R_2[y_0]W_1[y_1]W_2[x_1]C_1C_2$  & SER       \\
39     & $R_1[x_0]R_2[y_0]W_2[x_1]C_2W_1[y_1]C_1$  & SER        \\
40     & $R_1[x_0]R_2[y_0]W_1[y_1]C_1W_2[x_1]C_2$  & SER       \\ 
\hline
\end{tabular}
\label{tab:newpattern}
\end{table}

The anomaly pattern-guided transaction generation consists two steps. First we extract three types of constraints from each anomaly pattern, and then we generate transactions with the given database and SQL statements, guarded by those constraints. 
Given the lost update pattern (\(R_1[x_0]W_2[x_1]C_2W_1[x_2]C_1\)) in Table~\ref{tab:common_pattern}, \tool{} will generate a test case as shown in Figure~\ref{fig:generated}. We will progressively introduce how this is completed in this section.

\subsubsection{Constraint Extraction}
To implement anomaly pattern-guided transaction generation, we systematically extract three categories of constraints from each formalized anomaly pattern: statement type constraints, data access constraints, and scheduling constraints. These constraints define the characteristics of test transactions across three dimensions:
\begin{itemize}
    \item Operation dimension: The statement type constraint enforces that each operation within the generated transaction adheres to a specified type (e.g., read, write, commit).
    \item Data dimension: The data access constraint ensures that the statements access shared or conflicting data items, thereby creating the potential for inter-transaction interference.
    \item Temporal dimension: The scheduling constraint dictates the relative execution order of transactions to reflect the temporal conditions required by the anomaly pattern.
\end{itemize} 
Collectively, these constraints encode the essential semantics of an anomaly pattern,  providing a principled foundation for generating high-quality test transactions capable of exposing concurrency bugs.

\begin{definition}[Statement Type Constraint]
For a given anomaly pattern, the statement type constraint for a transaction is a four-tuple $<(R, n_R)$, $ (W, n_W)$, $(B, n_B)$, $(C, n_C)>$, which constraints on the number of read ($n_R$), write ($n_W$), rollback ($n_B$) and commit ($n_C$) statements in each generated transaction.
\end{definition}
For instance, given the lost update anomaly pattern $(R_1[x_0]W_2[x_1]$ $C_2W_1[x_2]C_1)$ in the example ~\ref{fig:generated},  
we can extract two statement type constraints $<(R, 1)$, $ (W, 1)$, $(B, 0)$, $(C, 0)>$ and $<(R, 0)$, $ (W, 1)$, $(B, 0)$, $(C, 1)>$ for the two involved transactions from the anomaly pattern through string parsing. 



\begin{definition}[Data Access Constraint]
The data access constraint establishes a mapping between an operation and the shared data it operates on in a given transaction. Formally, the mapping can be represented as $DA: \mathcal{O} \rightarrow V$, where $\mathcal{O}$ is the set of database operations R/W with inna-transaction timestamp, and $V$ is the set of shared variables in the given database. 
\end{definition}

For instance, given the lost update anomaly pattern ($(R_1[x_0]W_2[x_1]$ $C_2W_1[x_2]C_1)$), we can establish the following data access constraints: $DA(R_1^1)=x_0$, $DA(W_1^2)=x_2$ and $DA(W_2^1)=x_1$. 
The lower case of an operation represents the transaction id and the upper case represents the inna-transaction timestamp. In this example, $R_1^2$ represents that this is the second operation within transaction 1.   

\begin{definition}[Schedule Order Constraint]
The schedule order constraint regulates on the scheduling order of the generated concurrent transactions. It is defined as an integer sequence $[t_1, ..., t_m]$, where m is the total number of SQL statements in all concurrent transactions, and $t_i \in [1, n]$ is the transaction ID, with n being the number of concurrent transactions.  
\end{definition}
Taking again the lost update anomaly pattern as an example, the schedule order constraint extracted from this pattern is $[1,2,2,1,1]$. 


\begin{figure}[t] 
\centering 
\begin{lstlisting}
/*txn 1*/ START TRANSACTION;
/*txn 2*/ START TRANSACTION;
/*txn 1*/ SELECT * FROM t;
/*txn 2*/ UPDATE t SET c0 = 900 WHERE ID = 1;
/*txn 2*/ COMMIT;
/*txn 1*/ UPDATE t SET c0 = 1100 WHERE ID = 1;
/*txn 1*/ COMMIT;
\end{lstlisting}
\caption{A generated transaction guided by the lost update pattern (\(R_1[x_0]W_2[x_1]C_2W_1[x_2]C_1\))} 
\label{fig:generated} 
\end{figure}

These three types of constraints—statement type, data access, and schedule order—together form a semantically equipollent constraint set with the given anomaly pattern. In this way, we decompose the complex anomaly patterns into a set of simple constraints, which simplify our algorithm for anomaly pattern-guided transaction generation, which will be introduced in section~\ref{sec:transgen}.   

\begin{algorithm}[t]
\caption{Transaction Generation}
\label{alg:txn_case_gen}
\KwIn{$Cstr1$: statement type constraints, $Cstr2$: data access constraints, $Cstr_3$:  schedule order constraint}
\KwIn{$statements$, $database$}
\KwOut{Generated transactions}

AllStmts $\gets$ []

\For{tuple in $Cstr1$}{ 
    stmts $\gets$ getStatements($statements$, tuple) \TriComment{statement selection}
    
    AllStmts.add(stmts) 
}

conditions $\gets$ genConditions($database$, $Cstr2$) \TriComment{Generate conditions for variables}

AlignStmts $\gets$ init()

\For{op, var in $Cstr2$}{

    txnid $\gets$ getTxnid (op)
    
    inidx $\gets$ getInidx (op)
    
    stmt $\gets$ getStmt(stmts, op)
    
    cond $\gets$ getCond(conditions, var)
    
    newstmt $\gets$ AlignCondition(stmt, cond)  \TriComment{Condition Alignment}

    AlignStmts[txnid][inidx] $\gets$ newstmt 
}

transactions $gets$ []

\For{txnid in $Cstr3$}{

    stmt $\gets$ AlignStmts[txnid][0]

    AlignStmts[txnid].remove(stmt) 

    transactions.add(stmt)  \TriComment{Transaction Compose}
}

transaction.package()  \TriComment{Add BEGIN}

\end{algorithm}

\subsubsection{Transaction Generation. }
\label{sec:transgen}
The anomaly pattern-guided transaction generation method is shown in Algorithm~\ref{alg:txn_case_gen}. It takes the three types of constraints extracted from an anomaly pattern, the set of SQL statements and database generated by SQLancer as inputs and progressively construct transactions satisfying the constraints. 
Guided by the constraints, there are three main steps in the algorithm: statement selection, condition alignment, and transaction composition. 

\vspace{1mm}
\noindent\textbf{Statement Selection.} In this step, we select the appropriate SQL statement for each operation extracted from the pattern (lines 2-4). Based on the operation type (e.g., read, write, commit, rollback), the corresponding SQL statement is chosen from the set of SQL statements we randomly generated.  As shown in Figure~\ref{fig:generated}, we select SQL statements from lines 3 to 7 in the example. However, at this step, there is no guarantee on the exact WHERE conditions in the statements. For example, the selected statement may be like \textit{UPDATE t SET c0 = 900 WHERE c0 < 0} for line 4. 

\vspace{1mm}
\noindent\textbf{Condition Alignment.} In this step, we first generate the conditions corresponding to the variables in the data access constraints (line 5).  
Using data, a heuristic algorithm assigns a higher probability to the most recently inserted rows, ensuring comprehensive data coverage. Next, based on the table's attributes, we select the corresponding columns and generate conditions according to the syntax rules (line 5), such as \textit{ID = 1}. We then extract the relevant variables and conditions for the statement (lines 8-11). These conditions replace the initial ones in the statement (line 12). For example, the initial statement \texttt{UPDATE t SET c0 = 900 WHERE c0 < 0} is modified to \texttt{UPDATE t SET c0 = 900 WHERE ID = 1}. through coditio alignment.  Additionally, we allow the use of empty conditions, which represent a full table scan. Subsequently, we generate statements like those in lines 3, 4, and 6 in Example~\ref{fig:generated}. This approach ensures that statements involving the same variable operate on the same data items, thereby establishing potential dependencies that align with the pattern requirements. 

\vspace{1mm}
\noindent\textbf{Transaction Composition.}
In this step, based on the transaction schedule, we arrange the SQL statements in the correct execution order to ensure that the transaction follows the intended anomaly pattern (lines 15-18), similar to the execution order shown in Example~\ref{fig:generated}. Finally, we add the "BEGIN," "START TRANSACTION," or "START TRANSACTION WITH CONSISTENT SNAPSHOT" statements at the beginning of the transaction (line 19).

\begin{algorithm}[t]
\caption{Transaction Execution Schedule}
\label{alg:tx-scheduling}
\KwIn{$schedule$: Sequence of transaction ids}
\KwIn{$transactions$: generated transaction case}
$length \gets$ schedule.length()

$stmtStates \gets$ getStatements($transactions$)

\While{$\sum(\textit{stmtsStates.done}) < length$ \textbf{and} $Message = \varnothing$}{
    
    \For{$i = 0$ \KwTo $length - 1$}{
        $curTxn \gets$ getTxn(schedule[$i$])
        
        \If{$stmtStates[i].done$ \textbf{or} $curTxn.blocked$}{
            \textbf{continue} \TriComment{skip executed or blocked txn}
        }

        \underline{$curTxn.execute(stmtStates[i])$}

        \If{$curTxn.blocked$}{
            $waitTxn.add(curTxn)$ \TriComment{move to wait list}
            
            \textbf{continue}
        }

        \If{$curTxn.Message \neq \varnothing$}{
            $Message \gets curTxn.Message$
            
            \textbf{break}
        }

        $res \gets$ getRes($curTxn$) \TriComment{collect execution result}

        \ForEach{$txn$ \textbf{in} $waitTxn$}{
            \If{not $txn.blocked$}{
                $res \gets$ getRes($txn$) \TriComment{resume unblocked txn}
                
                $waitTxn.remove(txn)$
            }
        }
    }
}
\end{algorithm}

Finally, we generate a transaction case, as shown in Figure~\ref{fig:generated}, which can be executed after creating the database and setting the isolation level during the execution phase. The composed transaction is then ready for testing, providing a comprehensive test case for evaluating the database. 


\subsection{Execution}
\label{sec:execution} 
In the execution phase, our goal is to ensure that each transaction is executed in a predefined scheduling order, with the execution status and results of each transaction being accurately and efficiently recorded for bug detection. The key to this process is maintaining the correct execution order of the transactions while simultaneously recording the actual execution results. Additionally, we handle potential issues such as transaction blocking and rollback to ensure the correctness and reliability of the testing process. 


\subsubsection{Transaction Schedule} 
During execution, we must ensure that transactions are executed in the specified scheduling order, adhering to the scheduling policy established earlier. However, the database itself may handle exceptional situations through blocking and error mechanisms. Therefore, while maintaining the scheduling order, we must also ensure that test cases are not indefinitely blocked and avoid prematurely terminating test cases, which could result in a low number of executed test cases. Additionally, test cases that do not strictly follow a given the scheduled order may still trigger bugs. Therefore, we explicitly monitor system status such as transaction blocking and rollback in the execution phase. 

Algorithm~\ref{alg:tx-scheduling} demonstrates our approach to transaction scheduling, which incorporates strict control over the order of transaction execution while accounting for potential blocking and error mechanisms.In this algorithm, each transaction, represented by \textit{curTxn}, corresponds to a transaction thread responsible for executing SQL statements and storing the execution results and status.  
SQL statements are executed sequentially within their respective transactions according to the scheduling order (line 8). If a statement has already been executed or the current transaction is blocked, the statement is skipped (lines 6-7). When a transaction is blocked, it is added to the appropriate wait queue, and the current blocking status is recorded (lines 9-11). If the current statement can be executed normally, record the database message and execution result of the current statement execution (lines 12-15). 
Once the transaction's blocking is resolved (e.g., the previous transaction has committed, or a lock/resource has been released), the transaction is rescheduled for execution (lines 16-19). This mechanism simulates the complexity of real-world transaction scheduling and concurrency control, ensuring that the tests cover all potential concurrency conflicts. When all statements have been executed or an exceptional message occurs, the program is terminated.

Through this transaction scheduling mechanism, we can ensure that each transaction follows the predetermined execution order according to the pattern, while also allowing for database blocking and scheduling mechanisms. This provides correctness guarantee for transaction bug detection based on the recorded information. 

\subsubsection{Results Recording} 
During transaction execution, it is crucial to record the intermediate execution results of each transaction for subsequent bug detection and analysis. For read operations, the database typically returns the results directly, requiring no additional processing. However, for write operations, the database does not directly return the modified rows and data values, which presents a challenge in recording the results. To ensure that the results of write operations are fully documented, 
we use database triggers to capture the execution and results of write operations.
A trigger is a special type of stored procedure that is automatically executed in response to a specific event. We use triggers to capture data write operations, particularly for update, insert, and delete operations. By recording the data values before and after the write operation within the trigger, we can obtain a comprehensive view of the results of each operation and log them for subsequent bug detection and debugging.

\begin{figure}[t] 
\centering 
\begin{lstlisting}
CREATE TRIGGER tri_update_{table}
AFTER UPDATE ON {table}
FOR EACH ROW
BEGIN
    INSERT INTO {table}_log (operation_type, txn_id, row_id, version, time)
    VALUES ('UPDATE', CONNECTION_ID(), NEW.ID, NEW.VAL, SYSDATE(6));
END;
\end{lstlisting}

\caption{A MySQL trigger example of update statement} 
\label{fig:mysql_trigger} 
\end{figure}

Figure~\ref{fig:mysql_trigger} shows an example of trigger, where we log the execution values of the update before and after the operation. the logged information includes the operation type, the transaction ID, the updated data row ID, the data version and the timestamp indicating the actual execution order. 

We adopt triggers as means for recording the write execution results due to their simple implementation, real-time recording capability and strong adaptability to different RDBMSs and isolation levels. 
By using triggers, we are able to efficiently record the execution results of all write operations in a transaction in real time, ensuring that every modification is fully traceable. Triggers not only offer robust support for transaction bug detection by capturing detailed data change histories, but also significantly simplify the implementation of testing tools. This allows \tool{} to execute a wide variety of transaction workloads, increasing the likelihood of triggering and detecting bugs. 

\subsection{Detection} 
The detection phase is a fundamental component of the transaction testing framework, responsible for checking that each transactional operation complies with the expected semantics under the specified isolation level. \tool{} employs a two-phase detection strategy. In the first phase, it performs explicit error detection, capturing observable failures such as system crashes or assertion violations. If no explicit errors are observed, \tool{} proceeds to the second phase—implicit error detection—which analyzes execution results against a set of predefined anomaly patterns to uncover consistency violations. The implicit error detection phase is designed specifically to detect logic transaction bugs, with no explicit error messages. 


\subsubsection{Explicit Error Detection}
Explicit error detection refers to the direct detection of error messages—such as transaction crashes, assertion failures, and deadlocks—using the database system’s built-in mechanisms. Our detector monitors crash reports and assertion failures during execution, and immediately flags them as bugs when identified. 
Except for those error messages, we also monitor messages on deadlock, syntax error and semantic errors. The transactions raising those messages are stopped and discarded as they fail to provide valid execution traces, which in our cases refer to the traces leading to transaction processing code logic. We remove a large number of false positives through this filtering process, as is discussed in section~\ref{sec:exp-compare}.

\subsubsection{Implicit Error Detection}
Implicit error detection is a dynamic approach to error detection based on the real execution results of the database. It helps uncover errors that are not explicitly reported by the database system but may still compromise data consistency. \tool{}  constructs a dynamic execution dependency graph for pattern matching and verification, aiming to identify potential logical errors in transaction executions.
\begin{definition}[Dependency Graph]
A dependency graph is formally defined as $G=(N, E)$, where N is a set of nodes and E is the set of edges. Each node in G is a transaction, and each edge in G represents the dependencies, e.g., ww, wr, rw dependencies between transactions, as introduced in section~\ref{sec:dependency}. 
\end{definition}


For write-write (ww) dependencies, we check whether different transactions have written to the same record, using the unique ID column in the tables as the reference. Similarly, for read-write (rw) dependencies, we check whether a record read by one transaction was subsequently modified by another.  
The detection of write-read (wr) dependencies is more complex. It involves not only checking whether two transactions operate on the same records, but also checking whether the versions of those records match. Specifically, even if both the write and read operations target the same data item, a write-read dependency is only considered to exist if the read operation fails to observe the value written by the preceding write. In other words, the dependency is identified only when the expected data version is not reflected in the read, indicating a potential consistency violation.

When the dynamic dependency graph is constructed, we can use it to match predefined patterns. First, we extract a dependency path based on the anomaly pattern. For example, in the case of the lost update pattern, we identify an rw-ww dependency path. We then search the dependency graph to determine if such a dependency path exists. If found, we consider the match successful. When we successfully match a pattern that is disallowed by the isolation level, it indicates that an actual anomaly has occurred during execution. \tool{} then reports that a potential transaction bug has been triggered. 
\section{Experiment}
\label{sec:experiment}

\subsection{Experimental Setup}
\begin{table}[]
\centering
\caption{The tested RDBMSs and their supported isolation levels}
\label{tab:testedDB}
\begin{tabular}{ccccccc}
\hline
\multirow{2}{*}{DBMS} & \multirow{2}{*}{Stars} & \multirow{2}{*}{Loc} & \multicolumn{4}{c}{Isolation Level} \\ \cline{4-7} 
                      &                        &                      & RU      & RC      & RR     & SER     \\ \hline
MySQL                 & 11k                    & 6.0M                 & Yes     & Yes     & Yes    & Yes    \\
MariaDB               & 5.8k                   & 2.0M                 & Yes     & Yes     & Yes    & Yes    \\
OceanBase             & 8.7k                   & 7.1M                 & No      & Yes     & Yes    & Yes   \\ 
Postgres              & 19k                    & 1.5M                 & Yes     & Yes     & Yes    & Yes    \\
OpenGauss             & 1.5k                   & 21M                  & No      & Yes     & Yes    & No    \\ \hline
\end{tabular}
\end{table}


We conducted comprehensive experiments to evaluate the effectiveness and efficiency of \tool. All experiments were performed on a machine running Ubuntu 22.04, equipped with a 64-core Intel(R) Xeon(R) Gold 5218 CPU @ 2.30GHz and 160GB of RAM.

\vspace{1mm}
\noindent\textbf{RDBMSs Under Test.} In our evaluation, we examine three open-source RDBMSs—MySQL, MariaDB, and OceanBase (see \textit{Table~\ref{tab:testedDB}}), to demonstrate the effectiveness and generality of \tool. We also list the isolation levels supported by each database, all of which are testable using \tool. 
We tested the latest available versions of each database: MySQL 9.1 and 9.2, MariaDB 10.6.17 and 11.7.2, and OceanBase 4.2.1. All databases were deployed using Docker, with official images provided by the respective developers. 


We tested the read committed, repeatable read, and serializable isolation levels in MySQL, MariaDB, and OceanBase. The read uncommitted isolation level provides very weak guarantees and is rarely used in practice. Moreover, several databases, including OceanBase, do not support this isolation level. As a result, we excluded read uncommitted from our evaluation. 
Notably, while OceanBase claims that it offers the Serializable isolation level, its actual implementation is based on snapshot isolation, which allows write skew.  
%
For each isolation level of each database, we ran \tool{} continuously. We manually analyzed the bugs reported by \tool{} and submitted the identified issues to the corresponding development communities.

\vspace{1mm}
\noindent\textbf{Compared Approaches. }
\label{sec:exp-compare}
We compare \tool{} with two state-of-the-art testing methods: TxCheck~\cite{jiang2023detecting} and Troc~\cite{dou2023detecting}. TxCheck is a metamorphic testing approach, which first randomly generates test transactions and executes them. It then decouples the transaction into semantically equivalent SQL statements based on their dependency topology. After that, it compares the results of executing the decoupled statements with the results of the original transaction to identify inconsistencies. 
Troc is a differential testing approach, which decouples randomly generated transactions into independent SQL statements and constructs a view for each statement according to the semantics of its isolation level. It compares the execution result of the original transaction with the result derived from executing the individual statements within their respective views. This allows Troc to verify whethfer the overall transaction behavior is consistent with the isolation-level semantics of its component operations.


\subsection{Bug Detection Capability}
\begin{table}[t]
\centering
\small
\caption{Bugs detected by APTrans}
\label{tab:bug_list}
\begin{tabular}{m{1.2cm}m{0.8cm}m{1.5cm}m{1cm}m{1cm}m{0.8cm}}
\hline
\textbf{DBMS} &\textbf{Bug ID} &\textbf{Status} &\textbf{Severity} &\textbf{Isolation Level} &\textbf{Oracle} \\ \hline
\multirow{5}{=}{\centering MySQL} 
    & 115978 & Duplicate & Moderate & SER & Implicit \\
    & 117218 & Confirm & Moderate   & RR  & Implicit\\
    & 117733 & Confirm & Serious   & RC   & Implicit\\
    & 117835 & Confirm & Critical    & RR  & Implicit\\
    & 117860 & Confirm & Serious    & SER  & Implicit\\ \hline
    
\multirow{5}{=}{\centering Mariadb} 
    & 35335 & Confirm & Critical  & SER  & Implicit\\
    & 35464 & Confirm & Critical  & RR   & Implicit\\
    & 36120 & Inconsistency & Serious   & ALL  & Implicit \\
    & 36308 & Inconsistency & Critical  & RR   & Implicit\\
    & 36330 & Confirm & Critical  & SER  & Implicit\\
    \hline
\multirow{1}{=}{\centering OceanBase} 
    & 2248 & Confirm &  -    & RR   & Implicit\\ \hline

\multirow{1}{=}{\centering OpenGauss} 
    & IBZAAG & Confirm &  -    & ALL   & Explicit\\ \hline
\end{tabular}
\vspace{-4mm}
\end{table}


Over a seven-month period, \tool{} detected 13 unique transaction-related bugs across MySQL, MariaDB, and OceanBase. Of these, 11 have been confirmed by the respective development communities, and 10 were first reported by \tool. Detailed information of the detected bugs is presented in Table~\ref{tab:bug_list}.


Bugs were detected across all isolation levels evaluated in our study. Specifically, we identified one bug under the read committed (RC) isolation level, six bugs under the repeatable read (RR) isolation level, and six bugs under the serializable (SER) isolation level. 
While higher isolation levels are expected to offer stronger consistency guarantees, their low efficiency attracts RDBMS developers to implement optimization strategies to improve the transaction execution efficiency. Those optimization mechanisms potentially  
introduce subtle and difficult-to-detect implementation bugs. This finding highlights the critical importance of systematic testing in database transaction execution logic,  particularly under strong isolation levels where correctness is both essential and more challenging to achieve.

There are 10 out of 13 of the detected transaction errors classified as Critical or Serious, underscoring their significant impact on database correctness and reliability. Although the OceanBase community does not explicitly label bug severity, our analysis indicates that the issues found in OceanBase can also lead to serious data consistency violations. These high-severity errors highlight the importance of rigorous transaction testing—particularly under high isolation levels—where the concurrent execution of transactions can give rise to subtle, hard-to-detect bugs that compromise both data integrity and system stability.


All of the bugs were captured through the implicit error detection oracle. This emphasizes that many database transaction bugs do not trigger explicit error messages or system alerts, making them particularly difficult to detect using traditional error-handling methods. Consequently, implicit error detection methods are essential for uncovering these hidden issues. By analyzing transaction execution traces and conducting anomaly dependency checking, our approach can detect bugs even in the absence of explicit errors, ensuring a more comprehensive and reliable bug detection. 

For many of the bugs we discovered, we observed several well-known data anomalies—such as lost updates, dirty reads, and non-repeatable reads—that were not properly prevented by the tested database systems. These anomalies can lead to unpredictable and potentially severe errors in real-world applications. This observation underscores the effectiveness of anomaly-guided testing: instead of relying on large-scale random test case generation, we can systematically construct targeted test cases based on well-defined anomaly patterns, enabling more focused, efficient, and interpretable detection of transactional inconsistencies. 

\subsection{Comparison with Existing Approaches}
\begin{table}[t] 
\centering
\caption{Comparison of bug-finding capabilities in 72 hours}
\begin{tabular}{ccccc}
\hline
\textbf{Methods}         &\textbf{Category} & \textbf{MySQL} & \textbf{MariaDB} & \textbf{OceanBase}        \\ \hline
\multirow{2}{*}{\tool{}}    & Total            & 25            & 23              & 24                         \\ 
                         & True Positive    & 25            & 23              & 24                         \\ \hline
\multirow{2}{*}{TxCheck} & Total            & 24             & 73               & \diagbox{}{}                  \\ 
                         & True Positive    & 0              & 0                & \diagbox{}{}                  \\ \hline
\multirow{2}{*}{Troc}    & Total            & 23             & 18               & 39                        \\ 
                         & True Positive    & 0              & 0                & 0                         \\ \hline
\end{tabular}
\label{tab:comparison} 
\end{table}

We compare \tool{} with two state-of-the-art transaction bug testing approaches—Troc and TxCheck—on three databases: MySQL, MariaDB, and OceanBase. To ensure a fair comparison, all methods were executed on the same database versions—MySQL v8.0.28, MariaDB v10.8.3, and OceanBase v4.2.1BP10—for an equal duration of 72 hours. We report the number of bugs reported by each method during this period. Note that TxCheck does not support OceanBase, and therefore its results are not reported for that system. The experimental results are summarized in Table~\ref{tab:comparison}. 


\tool{} detected 72 bugs during three days' testing, all of which are true positives. 
The high true positive rate subjects to two main reasons.  
First, our definition of anomaly patterns (in Table~\ref{tab:newpattern}) is strict and precise. When constructing these patterns, we thoroughly analyze the various concurrency issues that may arise from database transactions under different isolation levels. We ensure that each anomaly pattern has clearly defined triggering conditions, which can effectively distinguish between different types of anomalies.  
Secondly, our two-phase detection process improves accuracy. In the first phase, we filter out error messages, e.g., syntax errors, deadlocks, that are unrelated to transaction isolation, thereby avoiding interference with the transaction analysis. In the second phase, we perform analysis based on the actual transaction execution traces, constructing a transaction dependency graph for pattern matching. When an anomaly pattern is matched, it indicates that the bug occurred during the actual execution process. This ensures that anomalies are only detected when the actual database behavior fully aligns with the predefined pattern during testing. 
For example, when testing for dirty reads under the RR isolation level, a dirty read anomaly is recognized only if transaction 1 reads a modification made by transaction 2 that has not yet been committed, and the modification is later rolled back. We strictly adhere to these criteria and classify a behavior as a bug only when all conditions are met. 
In this way, our method accurately identifies bugs in database transactions, ensuring that irrelevant behavior is not misclassified as an bug.

Troc reported a total of 80 bugs, all of which are false positives. Based on its bug detection mechanism, we categorize these false positives into three reasons. More than half (56 cases) of the false positives are introduced due to inconsistent implementation between Troc and the RDBMS being tested, among which 41 cases are due to snapshot creation timing mismatch and 15 cases are due to lock wait time differences. The second most common reason (16 cases) that introduces false positives is incorrect isolation level interpretation. For instance, it incorrectly assumes that Repeatable Read (RR) prevents phantom reads, which is not guaranteed according to the SQL standard. The remaining 8 cases are due to the result comparison algorithm, which is supposed to conduct bag  comparison rather than list comparison according to the SQL semantics. 
%
The false positives in Troc’s results are primarily attributed to its re-implementation of transaction execution logic. This re-implementation deviates from the actual behavior of the RDBMSs under tested. This highlights the deficiencies of differential testing approaches, i.e., the design choice and correctness of the underlying implementations have large impact on the execution results, resulting in high false positive rate. 


TxCheck reported a total of 97 bugs, all of which were false positives. 
These false positives arise from three main reasons. First, TxCheck is unable to handle cyclic dependencies in the transaction scheduling, leading to 46 false positives. Second, 35 cases result from incorrect dependency order analysis in complex subquery statements. The remaining 16 cases syntax/semantic errors in the randomly generated SQL statements, and are not transaction-related bugs.
%
The false positives in TxCheck arise from its inability to handle cyclic dependencies in dynamic scheduling and misjudgments in query dependencies, particularly with complex nested subqueries. These limitations are due to the method’s reliance on a simplified transaction oracle design, which fails to accurately capture real-world database behaviors, such as cyclic dependencies and intricate execution order interactions, ultimately leading to incorrect results.

\subsection{Ablation Study}
We evaluate the effectiveness of the main components of the proposed method through ablation studies. Specifically, we assess the contribution of two critical components: the pattern-guided transaction generation module and the implicit error-checking module we implemented. By removing these components in separate experiments, we can better understand their individual contributions to the system's overall performance. To ensure a fair comparison, all experiments are conducted on the same database versions—MySQL v9.2, MariaDB v11.7.2, and OceanBase 4.2.1—over a 24-hour period. The experimental results are presented in Table~\ref{tab:ablation}.

\begin{table}[t]
\centering
\caption{The ablation experiment results (24 hours)}
\label{tab:ablation}
\begin{tabular}{cccc}
\hline
\textbf{Methods}                                    & \textbf{MySQL}    & \textbf{MariaDB}  & \textbf{OceanBase}    \\ \hline
\multirow{1}{*}{\tool}                              & 9                 & 3                 & 4                     \\ 
\multirow{1}{*}{\tool-Pattern }                        & 4                 & 2                 & 2                     \\
\multirow{1}{*}{\tool-Implicit Checker}                & 0                 & 0                 & 0                      \\ \hline

\end{tabular} 
\end{table}

In Table~\ref{tab:ablation}, \texttt{\tool{}-Pattern} represents the method that replaces pattern-guided transaction generation with randomized transaction generation, where both SQL statements and the transaction scheduling are generated randomly. This simulation of a more generic, non-guided approach resulted in a noticeable decrease in the number of bugs reported. The outcome suggests that the randomized approach is significantly less efficient than the pattern-guided transaction generation method, triggering 8 less bugs in the 1-day experiment. The pattern-guided transaction generation offers a structured, informed way to generate transactions, enabling the triggering of a broader range of bugs with greater efficiency. This highlights the value of incorporating anomaly patterns into test case generation, as opposed to relying on random methods.

\texttt{\tool{}-Implicit Checker} represents the method of removing the implicit error detection module, we retained only the explicit error detection module. During a 1-day experiment, no bugs were detected with only explicit error detection. This outcome highlights that the majority of database transaction bugs are silent, and do not trigger obvious error messages during execution. This emphasizes a critical aspect of transaction-related bugs: they often occur without visible indicators, making them difficult to detect using traditional error-handling methods. The absence of detected bugs in this case underscores the need for carefully designed approaches that monitors database status, such as our implicit error-checking module, to uncover silent transaction bugs accurately.  

The ablation studies highlight the significant contributions of both the pattern-guide transaction generation, which shows better bug-triggering capability, and the implicit error detection module, which shows accurate bug detection capability. The Pattern-guided test case generation and implicit error-checking modules together improve the detection of a broader range of transaction-related bugs. These findings underscore the importance of combining structured test case generation with specialized error detection techniques to achieve more effective and comprehensive bug identification in database systems.

\begin{figure}[t] 
\centering 
\begin{lstlisting}
/*init*/ CREATE TABLE tYv10enE (c0 INT PRIMARY KEY);
/*init*/ INSERT INTO tYv10enE (c0) VALUES (10989748), (-1404643822);
/*txn 1*/ BEGIN;
/*txn 1*/ SET SESSION TRANSACTION ISOLATION LEVEL READ COMMITTED;
/*txn 2*/ BEGIN;
/*txn 2*/ SET SESSION TRANSACTION ISOLATION LEVEL READ COMMITTED;
/*txn 1*/ BEGIN;
/*txn 2*/ BEGIN;
/*txn 1*/ UPDATE tYv10enE SET c0 = 0 WHERE c0 = 10989748;
/*txn 2*/ SELECT * FROM tYv10enE WHERE (c0 >= 0); @-[[0]]@
/*txn 1*/ ROLLBACK;
/*txn 2*/ COMMIT;
\end{lstlisting}
\caption{MySQL Bug\#117835 reported at Read Committed} 
\label{fig:mysql-117835} 
\end{figure}
\subsection{Case Study}

In this section, we provide detailed discussion on several representative transaction bugs detected by \tool{}, analyzing the reasons behind their accurate detection. Through these case studies, we demonstrate how \tool{} effectively identifies and uncovers potential bugs related to transaction isolation levels in a real database environment.

\vspace{1mm}
\noindent \textbf{Case Study 1. } Figure~\ref{fig:mysql-117835} illustrates a dirty read bug in MySQL under the Read Committed isolation level. According to the definition of the Read Committed isolation level, it requires that a transaction cannot read uncommitted data, which corresponds to the dirty read anomaly.In this example, both transactions explicitly set the isolation level to Read Committed after starting (lines 3–6). Transaction 1 modifies a data item by setting its value to 0 (line 9). Before Transaction 1 commits, Transaction 2 reads the uncommitted value 0 (line 10). Subsequently, Transaction 1 rolls back its changes (line 11).

This leads to a dirty read anomaly, as transaction 2 observes a value that was uncommitted. This behavior follows the dirty read pattern \(W_1[x_1]R_2[x_1]A_1C_2\) as defined in Table~\ref{tab:common_pattern}. In this pattern, the reader (transaction 2) reads a value written by another transaction (transaction 1) that is subsequently rolled back. A dirty read can cause significant issues in scenarios requiring strict data consistency, such as financial transactions or inventory management systems, where the integrity of read values is crucial. 

This test case can be generated by \tool{} guided by the dirty read pattern in Table~\ref{tab:common_pattern}. During execution, \tool{} records the necessary execution logs and then it analyze the results and identify any dependencies. Upon analyzing the results, we find a write-read dependency edge between the write operation in line 9 and the read operation in line 10, where transaction 2 reads a write from transaction 1. Based on this dependency, we construct the dependency graph and observe that it matches the dirty read pattern. As a result, our method reports the bug.

\begin{figure}[t] 
\centering 
\begin{lstlisting}
/*init*/ CREATE TABLE t (ID INT, c0 INT);
/*init*/ INSERT INTO t VALUES (1, 1), (2, 2);
/*txn 1*/ SET SESSION TRANSACTION ISOLATION LEVEL REPEATABLE READ;
/*txn 1*/ START TRANSACTION WITH CONSISTENT SNAPSHOT;
/*txn 2*/ SET SESSION TRANSACTION ISOLATION LEVEL REPEATABLE READ;
/*txn 2*/ START TRANSACTION WITH CONSISTENT SNAPSHOT;
/*txn 1*/ UPDATE t SET c0 = 10 WHERE (ID = 1);
/*txn 1*/ COMMIT;
/*txn 2*/ SELECT * FROM t; @-[(1, 10), (2, 2)]@
/*txn 2*/ COMMIT;
\end{lstlisting}
\caption{OceanBase Bug\#2248 reported at Repeatable Read} 
\label{fig:ob-2248} 
\end{figure}

\vspace{1mm}
\noindent\textbf{Case Study 2. } Figure~\ref{fig:ob-2248} illustrates a non-repeatable read bug that occurs under the repeatable read isolation level in OceanBase. According to the definition of the repeatable read isolation level, it requires that the same transaction reads from the same snapshot; otherwise, a non-repeatable read anomaly occurs. In this example, two transactions both operate in the repeatable read isolation level, and begin with consistent snapshots (lines 3-6). In this example, the semantic of ‘START TRANSACTION WITH CONSISTENT SNAPSHOT’ is equivalent to starting a transaction and executing a SELECT statement to create a snapshot. Subsequently, transaction 1 modifies a row and commits (lines 7-8), while transaction 2—despite starting before transaction 1 commits—subsequently reads the modified value (line 9).  

This leads to a non-repeatable read anomaly, as transaction 2 observes changes that were committed after its snapshot was taken. This behavior follows the pattern \(W_1[x_1]\)\(C_2\)\(R_2[x_1]\)\(C_1\), as defined in Table~\ref{tab:newpattern}. In this pattern, the reader (transaction 2) reads a value written by another transaction (transaction 1) under the repeatable read isolation level.

The transaction that triggered this non-repeatable read bug was generated by \tool, guided by pattern 3 in Table~\ref{tab:newpattern}. For bug detection, we observe a write-read dependency edge (lines 7–9) in the execution log, which matches the pattern \(W_1[x_1]\)\(C_2\)\(R_2[x_1]\)\(C_1\), indicating that the Repeatable Read isolation level has been violated.

\vspace{1mm}
\begin{figure}[t] 
\centering 
\begin{lstlisting}
/*init*/ CREATE TABLE t (c0 INT PRIMARY KEY AUTO_INCREMENT, c1 INT);
/*txn 1*/ SET SESSION TRANSACTION ISOLATION LEVEL SERIALIZABLE;
/*txn 2*/ SET SESSION TRANSACTION ISOLATION LEVEL SERIALIZABLE;
/*txn 1*/ BEGIN;
/*txn 2*/ BEGIN;
/*txn 1*/ INSERT INTO t(c1) VALUES(1);
/*txn 2*/ INSERT INTO t(c1) VALUES(2), (3);
/*txn 2*/ COMMIT;
/*txn 1*/ SELECT * FROM t; -- @[(1, 1), (2, 2), (3, 3)]@
/*txn 1*/ COMMIT;
\end{lstlisting}
\caption{MariaDB Bug\#36330 reported at Serializable} 
\label{fig:mariadb-36330} 
\end{figure}
\noindent \textbf{Case Study 3.} Figure~\ref{fig:mariadb-36330} presents a bug observed in MariaDB under the Serializable isolation level. According to the definition of the Serializable isolation level, it requires that the concurrent execution of transactions should be equivalent to the result of some serial execution of those transactions. In this example, a table with an auto-increment primary key is created (line 1), and two transactions concurrently insert data into the table. After transaction 2 commits, transaction 1 performs a query and unexpectedly retrieves three rows: [(1,1), (2,2), (3,3)].

This result violates the serializable isolation level, as it does not correspond to any valid serial execution of the two transactions. If transaction 1 had executed first, the query should have returned only a single row: [(1,1)]. Conversely, if transaction 2 had executed first, due to the behavior of auto-increment columns, the query should have returned three rows: [(1,2), (2,3), (3,1)]. The observed result matches neither of these valid serial outcomes, indicating a non-serializable interleaving. More critically, if these query results were used in subsequent write operations, this bug could lead to even more severe data inconsistencies. 

This example can be generated and detected by our pattern 36 \(W_1[x_1]\)\(W_2[y_1]\)\(C_2\)\(R_1[y_1]\)\(C_1\) in Table~\ref{tab:newpattern}, which represents a cyclical anomaly arising from the combination of write-write and write-read dependencies.
\section{Related Work}
\label{sec:relatedwork}
In this section, we discuss works that are closely related with our approach. In particular, we discuss existing research on RDBMS transaction bug testing, RDBMS transaction history validation and general RDBMS bug testing. 

\vspace{1mm}
\noindent\textbf{RDBMS Transactional Bug Testing Methods.}
Troc~\cite{dou2023detecting} and TxCheck~\cite{jiang2023detecting} are recent methods for detecting database transaction bugs. Both generate transactions by randomly composing SQL statements, making it challenging to trigger transaction bugs, which usually occur under  specific patterns. Troc is a differential testing approach, it compares concurrent transaction execution results against its constructed views to identify differences in query outcomes. However, the view-construction resembles re-implementation of the transaction execution logic, which incurs overhead. Moreover, due to different design and implementation choices of its view with the databases under test, this method often leads to high false positives, as discussed in section~\ref{sec:exp-compare}.  
TxCheck decouples transactions into independent and dependent statements, and then generates semantically equivalent variants by re-scheduling independent statements while preserving the order of dependent statements. Then the original transaction and its  semantically equivalent variant are executed and the results are compared to identify inconsistencies. Nevertheless, since the decoupling process is based on topological sorting, it struggles with cyclic dependencies, limiting its ability to detect certain cyclic anomalies. \tool{} uses the anomaly pattern as a guide for both transaction generation and bug detection, which improves the ability of bug triggering and the accuracy of bug detection. 
%

DBStorm~\cite{li2024dbstorm} is a testing workload generation method that injects read-write operation sequences on the same data within generated transaction loads to create anomalies. However, it only generates five categories of anomalies, overlooking many others, such as non-repeatable reads. Additionally, the injection of anomalies into the transaction load does not guarantee the required scheduling order to trigger these anomalies. In contrast, our method extends the anomaly pattern and ensures the required   transaction schedule for more comprehensive anomaly generation and detection.

DT2~\cite{cui2022differentially} is a differential testing approach that generates random test cases, executes them in different databases, and compares the results to identify inconsistencies. It mainly detects database compatibility issues, but suffers from a high false positive rate due to differences in database design choices and implementations. Our method, on the other hand, does not require comparison of execution results across multiple RDBMSs, which effectively reduces false positives and improves bug detection accuracy. 

\vspace{1mm}
\noindent\textbf{Transaction History Validation Methods.}
Transaction history validation~\cite{kingsbury2020elle, tan2020cobra, huang2023efficient, zhang2023viper, liu2024plume} aims to verify the correctness and integrity of database transaction sequences by analyzing transaction logs/execution histories generated during concurrent operations. Unlike testing, which focuses on generating transactions to actively trigger and then identify potential bugs, transaction history validation conducts post-execution analysis of transaction outcomes passively.

Cobra ~\cite{tan2020cobra} provides a highly scalable framework employing an SMT solver to address the computational explosion problem when detecting serializability violations in databases based on transaction history records. However, it suffers from lower performance under specific workloads and does not support advanced operations like native range queries. 
%
Elle~\cite{kingsbury2020elle} identifies isolation bugs by analyzing the transaction execution history on specifically designed consistency models (e.g., AppendList). It has the advantages of handling multi-type transaction patterns and efficiently locating faults. However, its reliance on characteristic models also limits its generality to some extent.
Recent optimizations include PolySI~\cite{huang2023efficient} and Viper~\cite{zhang2023viper}, which focus on improvements for Snapshot Isolation (SI) validation. Plume~\cite{liu2024plume}, which uses vector clock-based dependency graph analysis for weak isolation levels. However, these methods remain largely limited to key-value data models, lacking comprehensive support for SQL transaction semantics, such as multi-row operations in single statements, constraint validations, and complex data structures. These approaches focus on analyzing bugs from existing execution histories, rather than generating test cases to proactively test the database as our method does.

\vspace{1mm}
\noindent\textbf{General RDBMS Testing Methods. }
There are several general relational DBMS testing methods focusing on the execution behavior of SQL statements. SQLSmith~\cite{SQLsmith}, a pioneering generation-based approach, detects crashes by generating random SQL queries and monitoring for explicit errors. Its successor, SQLancer~\cite{rigger2020pqs, rigger2020norec, rigger2020tlp}, introduces three metamorphic testing oracles: PQS~\cite{rigger2020pqs}, NoREC~\cite{rigger2020norec}, and TLP~\cite{rigger2020tlp}, significantly improving its ability to detect logical bugs. Recent innovations have introduced paradigm shifts in testing strategies: QPG~\cite{ba2023testing} uses query plan-guided database state mutation, while DQE~\cite{song2023testing} applies differential execution analysis across SELECT, UPDATE, and DELETE statements. TQS~\cite{tang2023detecting} specifically targets join optimization errors with an oracle based on table partitioning. ETT~\cite{jiang2024detecting} detects bugs through equivalent expression transformation, and THANOS~\cite{fu2024thanos} identifies RDBMS bugs via differential testing, involving storage engine rotation.

There are also fuzzing approaches on RDBMSs. 
SQUIRREL~\cite{zhong2020squirrel} integrates language validity constraints with coverage feedback mechanisms, enhancing the generation of test cases. SQLRight~\cite{liang2022detecting} implements semantic-aware, coverage-guided mutation to improve the effectiveness of testing. GRIFFIN~\cite{fu2022griffin} advances metadata dependency analysis through statement reshuffling in metadata graphs, providing deeper insights into transaction behavior. 
AMOEBA~\cite{liu2022automatic} and APOLLO~\cite{jung2019apollo} concentrate on performance anomaly detection, utilizing workload pattern analysis and query plan comparison, respectively. 
Our work focuses specifically on detecting bugs in transactional systems, emphasizing the generation of concurrent transactions and the validation of correctness criteria that vary significantly across different isolation levels. This approach fundamentally differs from existing methods. 
\section{Conclusion}
\label{sec:conclusion}
We proposed a novel, general, and effective transaction testing approach, \tool, designed to accurately and efficiently detect transaction bugs in relational database management systems (RDBMSs). The core of \tool{} lies in generating test cases, i.e., concurrent transactions, which could potentially trigger transaction bugs and detecting those bugs accurately.
To achieve the above goal, we propose the anomaly pattern-guided transaction generation method, which can generate transactions satisfying the given anomaly pattern, increasing the likelihood of triggering bugs. We further propose a two-phase bug detection method, combining the ability of explicit bug detection with implicit bug detection based on carefully monitored database states.   
We applied \tool{} to test three widely used relational database systems—MySQL, MariaDB, and OceanBase—and identified 13 unique transaction bugs, among which 12 are reported for the first time and 11 are confirmed by the respective developers. The comparison results with two state-of-the-art transaction testing approaches show that \tool{} is superior on both transaction bug triggering and transaction bug detection accuracy.  



\balance
\bibliographystyle{ACM-Reference-Format}
\bibliography{bibtex}

\end{document}